\documentclass[a4paper,11pt]{article}
\pdfoutput=1 
\usepackage{aas_macros,braket}
\usepackage{jcappub,natbib} 
\usepackage[T1]{fontenc} 

\title{CMB bounds on tensor-scalar-scalar inflationary correlations}

\author[a]{Maresuke Shiraishi,}
\author[b,c,d]{Michele Liguori,}
\author[e]{and James R. Fergusson}

\affiliation[a]{Department of General Education, National Institute of Technology, Kagawa College, 355 Chokushi-cho, Takamatsu, Kagawa 761-8058, Japan}
\affiliation[b]{Dipartimento di Fisica e Astronomia ``G. Galilei'', Universit\`a degli Studi di Padova, via Marzolo 8, I-35131, Padova, Italy}
\affiliation[c]{INFN, Sezione di Padova, via Marzolo 8, I-35131, Padova, Italy}
\affiliation[d]{INAF-Osservatorio Astronomico di Padova, Vicolo dell'OSservatorio 5, I-35122 Padova, Italy}
\affiliation[e]{Centre for Theoretical Cosmology, DAMTP, University of Cambridge, Wilberforce Road, Cambridge, CB3 0WA United Kingdom}

\emailAdd{shiraishi-m@t.kagawa-nct.ac.jp}
\emailAdd{michele.liguori@pd.infn.it}
\emailAdd{J.Fergusson@damtp.cam.ac.uk}

\abstract{
  The nonlinear interaction between one graviton and two scalars is enhanced in specific inflationary models, potentially leading to distinguishable signatures in the bispectrum of the cosmic microwave background (CMB) anisotropies. We develop the tools to examine such bispectrum signatures, and show a first application using WMAP temperature data. We consider several $\ell$-ranges, estimating the $g_{tss}$ amplitude parameter, by means of the so-called separable modal methodology. We do not find any evidence of a tensor-scalar-scalar signal at any scale. Our tightest bound on the size of the tensor-scalar-scalar correlator is derived from our measurement including all the multipoles in the range $ 2 \leq \ell \leq 500$ and it reads $g_{tss} = -48 \pm 28$~($68\%$CL). This is the first direct observational constraint on the primordial tensor-scalar-scalar correlation, and it will be cross-checked and improved by applying the same pipeline to high-resolution temperature and polarization data from {\it Planck} and forthcoming CMB experiments.  
}

\begin{document}



\maketitle
\flushbottom

\section{Introduction}

Gaussianity of primordial density fluctuations is a natural prediction of simple early Universe scenarios such as a single-field slow-roll inflation based on Einstein gravity \cite{Acquaviva:2002ud,Maldacena:2002vr}. Conversely, a departure from Gaussianity indicates other, nonsimple scenarios, including e.g. the existence of multiple fields, nontrivial particle productions, the modification of Einstein gravity, and so on (see refs.~\cite{Bartolo:2004if,Liguori:2010hx,Chen:2010xka,Komatsu:2010hc,Yadav:2010fz} for some reviews). 

In light of this, primordial non-Gaussianity (NG) is a key inflationary indicator and therefore has been widely and deeply investigated from both the theoretical and the observational side. NG features, generated at primordial stages, are directly imprinted on late-time observables such as the cosmic microwave background (CMB) fluctuations, galaxy clustering, 21-cm anisotropies, and so on. 

Among these observables, we focus here on the CMB. A large set of NG primordial scenarios has already been tested with CMB data \cite{Bennett:2012zja,Ade:2013ydc,Ade:2015ava}. Recently, the {\it Planck} team has produced the most stringent constraints to date on the main NG inflationary shapes, namely local, equilateral and orthogonal (LEO), using both temperature and E-mode polarization data \cite{Ade:2015ava}. Such constraints, derived from the CMB bispectrum, indicate no evidence of these NG signatures at $> 2 \sigma$ level \cite{Ade:2015ava}. {\it Planck} NG studies for these shapes have also been extended to  CMB trispectra such of the $g_{\rm NL}$ and $\tau_{\rm NL}$-type NGs. Moreover, a vast amount of additional models, beyond LEO shapes, have been considered, finding data consistent with Gaussianity in all cases. Interestingly, some evidence of oscillatory bispectra at {\em specific} frequencies was found in the data \cite{Fergusson:2014hya,Fergusson:2014tza,Ade:2015ava}. However, the statistical significance of such signals vanish when accounting for ``look elsewhere" effects, due to the fact that the analysis is blindly scanning over a large range of frequencies.

The analysis mentioned above is thorough and includes a very large number of models. Nonetheless, it is mostly based on the auto-bispectrum of scalar modes, with some analysis of tensor bispectra \cite{Shiraishi:2014ila,Ade:2015ava}. Differently from the scalar case, tensor NG also generates CMB bispectra including B-mode polarization \cite{Shiraishi:2011st,Shiraishi:2013vha,Shiraishi:2013kxa,Shiraishi:2016yun,Tahara:2017wud}. Moreover, nonvanishing signals can arise also for odd $\ell_1 + \ell_2 + \ell_3$ configurations in the temperature or E-mode bispectra. This is sourced by parity violation in the tensor sector \cite{Kamionkowski:2010rb,Shiraishi:2011st,Shiraishi:2012sn,Shiraishi:2013kxa,Shiraishi:2014roa}.

Interestingly, nonlinear dynamics during inflationary stages can however also produce nonvanishing {\em scalar-tensor} couplings. Depending on the model under study, scalar-tensor signals can be amplified, up to detectable levels \cite{Domenech:2017kno}.%
\footnote{See refs.~\cite{Namba:2015gja,Agrawal:2017awz} for the inflationary models realizing sizable tensor auto-bispectra.}
Such mixed bispectra are currently unconstrained and this motivates the present work, where we will develop a general pipeline for the tensor-scalar-scalar bispectrum estimation, considering an interesting bispectrum shape realized in massive gravity \cite{Domenech:2017kno}, and show its first application, using WMAP data \cite{Bennett:2012zja,Hinshaw:2012aka}. This model generates a squeezed-type CMB bispectrum and allows for potential high signal-to-noise even at WMAP resolution, see \cite{Shiraishi:2010kd,Domenech:2017kno}. In general, CMB shapes arising from tensor primordial bispectra display a complex, nonseparable $\ell$ dependence, which makes a brute-force estimation approach numerically unfeasible.  To deal with this issue, we will adopt the so-called separable modal decomposition technique \cite{Fergusson:2009nv,Fergusson:2010dm}. As we will discuss in the following, our pipeline was thoroughly validated using NG simulations, generated as part of the analysis. The size of the tensor-scalar-scalar coupling was then estimated for several $\ell$-ranges, to check the scale dependence of the constrained values. Although, at the end of the analysis, we do not find any significant signal, we obtain meaningful observational bounds on the tensor-scalar-scalar coupling.

This paper is organized as follows: Sec.~\ref{sec:theory} illustrates the primordial tensor-scalar-scalar primordial correlator constrained in this paper and its signature in the CMB temperature bispectrum, Sec.~\ref{sec:observation} discusses the analysis of WMAP data and derives experimental constraints on the model, and Sec.~\ref{sec:conclusions} concludes this paper with a summary of its main results.

\section{Temperature bispectrum from a tensor-scalar-scalar correlator}\label{sec:theory}

In this paper we constrain the cross-bispectrum of primordial tensor and curvature perturbations, parametrized as \cite{Maldacena:2002vr,Shiraishi:2010kd}
\begin{eqnarray}
\Braket{\gamma_{{\bf k}_1}^{(\lambda_1)} \zeta_{{\bf k}_2} \zeta_{{\bf k}_3}}
 &=&
(2 \pi)^3 \delta^{(3)}\left(\sum_{n=1}^3 {\bf k}_n\right) e_{ij}^{(-\lambda_1)}(\hat{k}_1) \hat{k}_{2i} \hat{k}_{3j} 
\frac{16 \pi^4 g_{tss} A_S^2 }{k_1^2 k_2^2 k_3^2} \frac{I_{k_1 k_2 k_3}}{k_1}~, \label{eq:hzeta2} 
\end{eqnarray}
where $A_S$ denotes the amplitude of the primordial scalar power spectrum, and 
\begin{eqnarray}
  I_{k_1 k_2 k_3} \equiv  -k_t + \frac{k_1 k_2 + k_2 k_3 + k_3
 k_1}{k_t} + \frac{k_1 k_2 k_3}{k_t^2} 
\end{eqnarray}
with $k_t \equiv k_1 + k_2 + k_3$. The polarization tensor $e_{ij}^{(\pm 2)}$, obeying $e_{ii}^{(\lambda)}(\hat{k}) = \hat{k}_i e_{ij}^{(\lambda)}(\hat{k}) = 0$, $e_{ij}^{(\lambda) *}(\hat{k}) = e_{ij}^{(-\lambda)}(\hat{k}) = e_{ij}^{(\lambda)}(- \hat{k})$ and $e_{ij}^{(\lambda)}(\hat{k}) e_{ij}^{(\lambda')}(\hat{k}) = 2 \delta_{\lambda, -\lambda'}$, is used for the decomposition of the primordial gravitational wave into the helicity ($\lambda = \pm 2$) basis as
\begin{eqnarray}
 \gamma_{ij}({\bf x}) \equiv \frac{\delta g_{ij}^{TT}}{a^2} = \int \frac{d^3 {\bf k}}{(2\pi)^3}
 e^{i {\bf k} \cdot {\bf x}} \sum_{\lambda = \pm 2} \gamma_{\bf k}^{(\lambda)} e_{ij}^{(\lambda)}(\hat{k}) ~.
\end{eqnarray}
This type of bispectrum is actually realized in the simplest, single-field slow-roll inflationary models, based on Einstein gravity. Its amplitude, in this case, is as usual  determined by slow-roll parameters, namely, $|g_{tss}| \simeq \epsilon$. This makes the signal undetectably small \cite{Maldacena:2002vr}. By contrast,  ref.~\cite{Domenech:2017kno} has recently shown that, by introducing nonzero mass of gravitons, a nontrivial nonlinear coupling is induced, resulting in a sizable enhancement of the tensor-scalar-scalar signal: $|g_{tss}| \simeq \epsilon \lambda_{sst}$, with $\lambda_{sst}$ characterizing the coupling strength.%
\footnote{This model actually induces an additional term in eq.~\eqref{eq:hzeta2}. However, such is suppressed by the slow-roll parameter, hence negligible in a phenomenological analysis \cite{Domenech:2017kno}.}

For large enough $\lambda_{sst}$, the enhancement can produce detectable signals using current CMB datasets. It was in fact shown that  {\it Planck} temperature and E-mode polarization data allow achieving a $g_{tss} \sim 1$ sensitivity \cite{Shiraishi:2010kd, Domenech:2017kno}, with a further an-order-of-magnitude improvement possible using B-mode information \cite{Meerburg:2016ecv,Domenech:2017kno}. In this paper we will develop a tensor-scalar-scalar CMB bispectrum estimation pipeline, test it and use it for a preliminary analysis of WMAP data, reaching a sensitivity level $g_{tss} \sim 10$.

\begin{figure}[t]
\begin{center}
    \includegraphics[width=1.\textwidth]{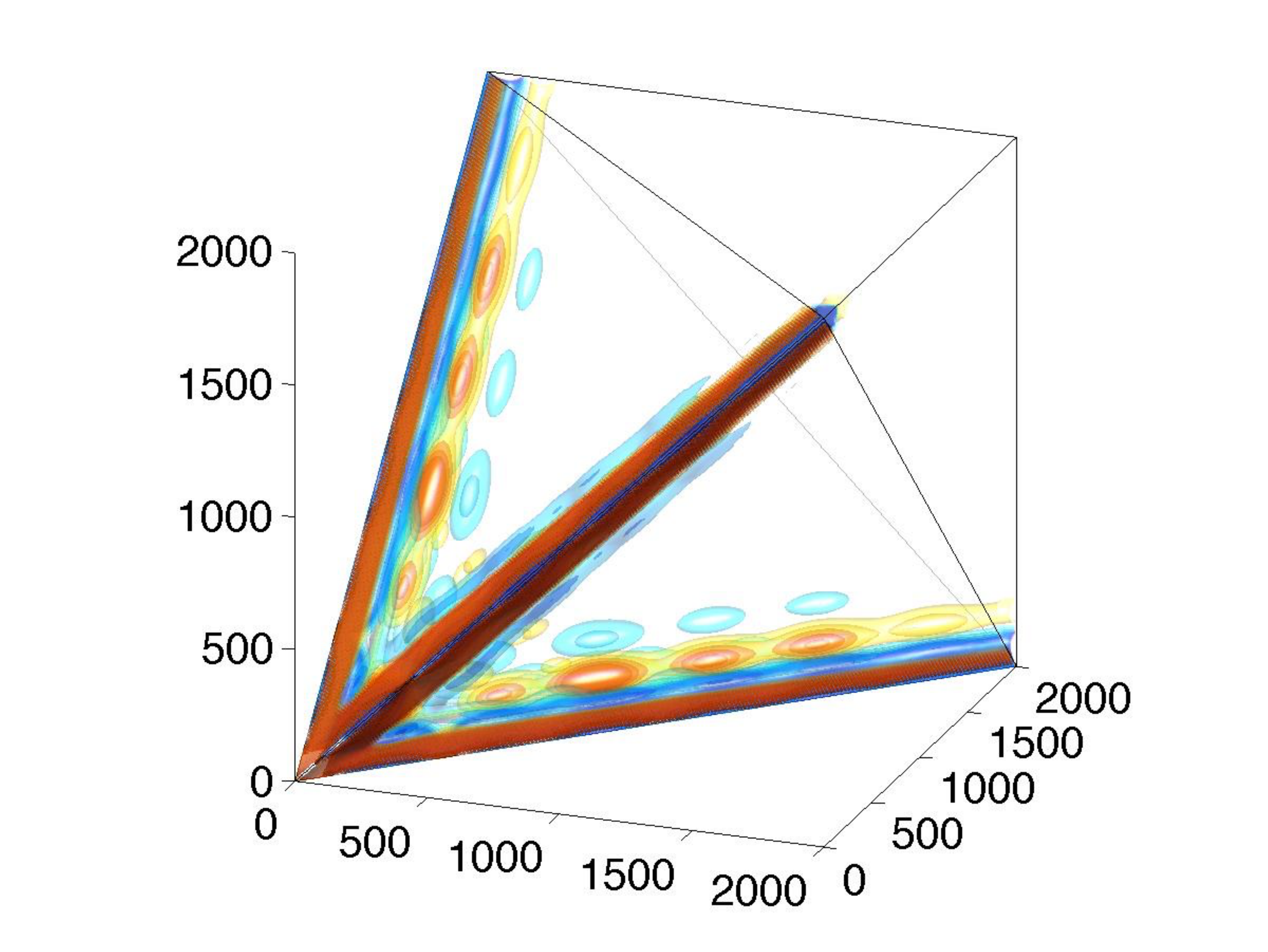}
\end{center}
    \caption{Three-dimensional representation of the CMB temperature bispectrum from the tensor-scalar-scalar correlator \eqref{eq:hzeta2} in the tetrahedral domain. We here plot $b_{\ell_1 \ell_2 \ell_3}$ normalized with a constant Sachs-Wolfe template \cite{Fergusson:2009nv} to highlight the dominant configurations. Three axes correspond to $\ell_1$, $\ell_2$ and $\ell_3$, respectively. Dense red (blue) color represents remarkable positive (negative) signal.} \label{fig:bis_3D}
\end{figure}

When the temperature fluctuation is multipole expanded in the usual manner, $T(\hat{n}) = \sum_{\ell m} a_{\ell m} Y_{\ell m}(\hat{n})$, the harmonic coefficients of the scalar and tensor modes are expressed, respectively, as
\begin{eqnarray}
  a_{\ell m}^{(s)} &=& 4\pi i^{\ell} \int \frac{d^3 {\bf k}}{(2\pi)^{3}} {\cal T}_{\ell}^{(s)}(k) \zeta_{\bf k}  Y_{\ell m}^*(\hat{k}) ~, \\
  a_{\ell m}^{(t)} &=& 4\pi i^{\ell} \int \frac{d^3 {\bf k}}{(2\pi)^{3}} {\cal T}_{\ell}^{(t)}(k)  \sum_{\lambda = \pm 2} \gamma_{\bf k}^{(\lambda)} {}_{-\lambda} Y_{\ell m}^*(\hat{k}) ~,
\end{eqnarray}
where ${\cal T}_{\ell}^{(s)}(k)$ (${\cal T}_{\ell}^{(t)}(k)$) is the radiation transfer function of the scalar (tensor) mode. According to ref.~\cite{Shiraishi:2010kd}, applying radiative transfer to eq.~\eqref{eq:hzeta2} leads to the CMB angular bispectrum, $\Braket{a_{\ell_1 m_1}^{(t)} a_{\ell_2 m_2}^{(s)} a_{\ell_3 m_3}^{(s)} }
  = g_{tss} B^{(tss)}_{\ell_1 \ell_2 \ell_3}
  \left(
  \begin{array}{ccc}
  \ell_1 & \ell_2 & \ell_3 \\
  m_1 & m_2 & m_3 
  \end{array}
 \right)$,
where 
    \begin{eqnarray}
  B^{(tss)}_{\ell_1 \ell_2 \ell_3}
  &=& \frac{(8 \pi)^{3/2}}{3} 
  \sum_{\substack{ L_1 = |\ell_1 \pm 2|, \ell_1 \\ L_2 = |\ell_2 \pm 1| \\ L_3 = |\ell_3 \pm 1|}}
  (-1)^{\sum_{n=1}^3 \frac{L_n + \ell_n}{2}} 
  h_{L_1 L_2 L_3} 
h^{2 0 -2}_{\ell_1 L_1 2} h_{\ell_2 L_2 1} h_{\ell_3 L_3 1} 
\left\{
  \begin{array}{ccc}
  \ell_1 & \ell_2 & \ell_3 \\
  L_1 & L_2 & L_3 \\
  2 & 1 & 1
  \end{array}
 \right\} \nonumber \\
&& \int_0^\infty y^2 dy  
\frac{2}{\pi}  \int_0^\infty k_1^2 d
  k_1 {\cal T}_{\ell_1}^{(t)}(k_1) j_{L_1}(k_1 y)
\left[ \prod_{n=2}^3  \frac{2}{\pi}  \int_0^\infty k_n^2 d
  k_n {\cal T}_{\ell_n}^{(s)}(k_n) j_{L_n}(k_n y)  \right] 
\nonumber \\
&& 
\frac{16 \pi^4  A_S^2 }{k_1^2 k_2^2 k_3^2} \frac{I_{k_1 k_2 k_3}}{k_1} ~,
\label{eq:CMB_bis_general}
\end{eqnarray}
with $h^{s_1 s_2 s_3}_{l_1 l_2 l_3}
\equiv \sqrt{\frac{(2 l_1 + 1)(2 l_2 + 1)(2 l_3 + 1)}{4 \pi}}
\left(
  \begin{array}{ccc}
  l_1 & l_2 & l_3 \\
  s_1 & s_2 & s_3
  \end{array}
  \right)$ and $h_{l_1 l_2 l_3} \equiv h^{0~0~0}_{l_1 l_2 l_3}$. Because of parity conservation and rotational symmetry in eq.~\eqref{eq:hzeta2}, nonvanishing signal is confined to the $\ell$-space domain satisfying $\ell_1 + \ell_2 + \ell_3 = {\rm even}$ and $|\ell_1 - \ell_2| \leq \ell_3 \leq \ell_1 + \ell_2$.

\begin{figure}[t]
\begin{center}
    \includegraphics[width=1.\textwidth]{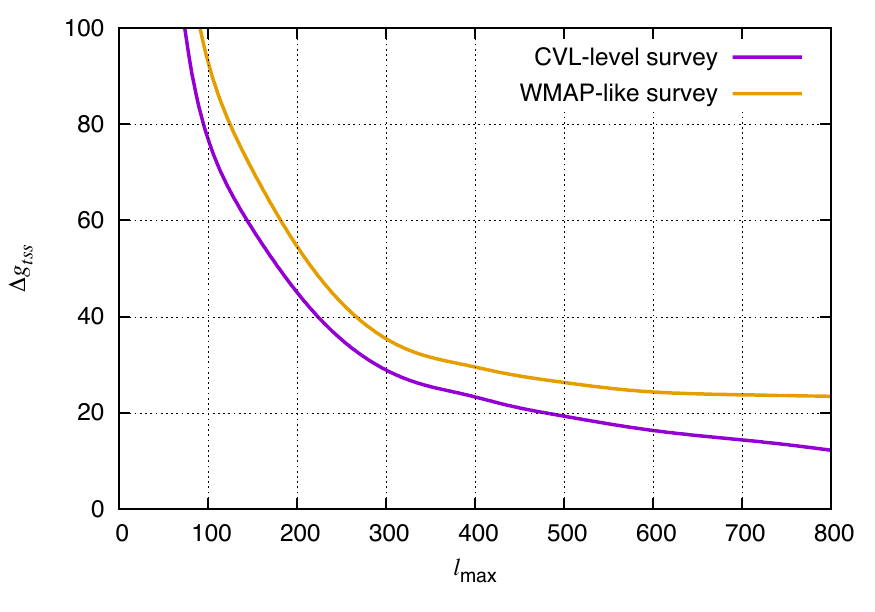}
\end{center}
    \caption{Fisher matrix forecasts of $1\sigma$ errors on $g_{tss}$ as a function of $\ell_{\rm max}$ assuming a CVL-level survey and the WMAP one.} \label{fig:error_WMAP}
\end{figure}

The observed bispectrum is given by the sum of all cyclic terms as $B_{\ell_1 \ell_2 \ell_3} = B^{(tss)}_{\ell_1 \ell_2 \ell_3} + B^{(sts)}_{\ell_1 \ell_2 \ell_3} + B^{(sst)}_{\ell_1 \ell_2 \ell_3}$, where $B^{(tss)}_{\ell_1 \ell_2 \ell_3} = B^{(sts)}_{\ell_3 \ell_1 \ell_2} = B^{(sst)}_{\ell_2 \ell_3 \ell_1}$. As it is customary in CMB bispectrum studies, let us introduce a ``reduced bispectrum'', $b_{\ell_1 \ell_2 \ell_3}$,  defined as $b_{\ell_1 \ell_2 \ell_3} = B_{\ell_1 \ell_2 \ell_3} / h_{\ell_1 \ell_2 \ell_3}$. Figure~\ref{fig:bis_3D} displays the shape of $b_{\ell_1 \ell_2 \ell_3}$ in $\ell$ space. It is clear that the shape under study strongly peaks on squeezed configurations (e.g. $\ell_1 \ll \ell_2 \sim \ell_3$). The signal-to-noise ratio grows in this case like $ \propto \ell_{\rm max}$, with $\ell_{\rm max}$ denoting the maximum available multipole \cite{Shiraishi:2010kd, Komatsu:2001rj}. Figure~\ref{fig:error_WMAP} shows expected $1\sigma$ errors, $\Delta g_{tss}$, from a Fisher matrix analysis, for an ideal, full-sky cosmic-variance-limited-level (CVL-level) measurement and for a realistic one, including WMAP-level instrumental noise. We see that for a WMAP-like survey, signal-to-noise essentially saturates at $\ell_{\rm max} \gtrsim 500$. Therefore we will discard multipoles with $\ell > 500$ in the data analysis phase, discussed below.

Because of the similarity between the two shapes (both peaking in the squeezed limit), one may think that the tensor-scalar-scalar bispectrum, discussed here, and the usual local-type shape, parametrized by $f_{\rm NL}^{\rm local}$, are completely degenerate. In fact this is not the case. The correlation coefficient between the two shapes falls rapidly for $\ell_{\rm max} \gtrsim 100$, due to the difference between their oscillating patterns \cite{Domenech:2017kno}. This allows to measure $g_{tss}$ independently of $f_{\rm NL}^{\rm local}$.

\section{CMB bispectrum estimations}\label{sec:observation}

In this section we measure $g_{tss}$ with the WMAP temperature data. We will assume that $g_{tss}$ is small so the NG signal is subdominant. We may therefore employ an optimal bispectrum estimator:
\begin{equation}
  \hat{g}_{tss} = \frac{1}{N}
  \sum_{\substack{\ell_1 \ell_2 \ell_3 \\ m_1 m_2 m_3}}
\left(
\begin{array}{ccc}
\ell_1 & \ell_2 & \ell_3 \\
m_1 & m_2 & m_3
\end{array}
\right)
h_{\ell_1 \ell_2 \ell_3} b_{\ell_1 \ell_2 \ell_3}
\left[ \prod_{n=1}^3
\frac{a_{\ell_n m_n}}{C_{\ell_n}} 
- 3 \frac{a_{\ell_1 m_1}}{C_{\ell_1}} 
\frac{\Braket{a_{\ell_2 m_2} a_{\ell_3 m_3}}_{\rm MC} }{C_{\ell_2} C_{\ell_3}}
 \right]~,
\label{eq:estimator}
\end{equation}
where $C_{\ell}$ is the temperature power spectrum and $N \equiv \sum_{\ell_1 \ell_2 \ell_3} (h_{\ell_1 \ell_2 \ell_3} b_{\ell_1 \ell_2 \ell_3})^2 / (C_{\ell_1} C_{\ell_2} C_{\ell_3})$ is a normalization factor. The input CMB coefficients $a_{\ell m}$ are computed from the observed data or simulated maps, and $\Braket{\cdots}_{\rm MC}$ is given by an ensemble average of the products of Gaussian simulated maps with realistic experimental features (i.e., beam shape, anisotropic noise and sky cut).
 
Bispectrum estimation using eq.~\eqref{eq:estimator} is fully optimal as long as the $signal$ + $noise$ covariance matrix is diagonal. This is of course not true in realistic cases, due to e.g. sky masking and nonstationary noise. To completely restore optimality, in this case, full-inverse covariance weighting of the data is required, rather than using just diagonal covariance elements, as in the formula above. While the numerical problem of producing inverse covariance weighted CMB maps at WMAP or {\it Planck} resolution is fully solved nowadays \cite{Smith:2007rg,Senatore:2009gt,Elsner:2012fe}, it has also been shown, for bispectrum analyses, that just a minimum ($\sim 5\%$) loss of optimality can be achieved by means of a much simpler, recursive inpainting pre-filtering technique \cite{Gruetjen:2015sta,Bucher:2015ura}. This is the approach adopted in {\it Planck} papers \cite{Ade:2013ydc,Ade:2015ava}, and followed also in the current analysis.

The best-fit value of $g_{tss}$ is determined from a coadded temperature map, using WMAP 9-years, V and W bands \cite{Bennett:2012zja,Hinshaw:2012aka}. The results reported in this paper are obtained from foreground-cleaned data, and we checked that (not-foreground-cleaned) raw data also gives compatible values. For Monte-Carlo evaluation of the error on $g_{tss}$ and to compute the linear correction term in the estimator ($\Braket{\cdots}_{\rm MC}$ in the formula above), we use 500 random Gaussian simulations and follow the WMAP-team methodology \cite{Komatsu:2008hk} to include the effects of anisotropic noise and beam shape in the relevant frequency bands. We mask both observations and simulations with the KQ75 mask, characterized by a sky coverage fraction $f_{\rm sky} = 0.688$. After removing monopole and dipole components, masked regions are inpainted via the recursive inpainting pre-filtering technique mentioned above. All the data and instrumental information used here are provided by the Lambda website~\cite{Lambda}.

\subsection{Modal decomposition}

Even if $b_{\ell_1 \ell_2 \ell_3}$, $a_{\ell m}$, $C_\ell$ and $\Braket{\cdots}_{\rm MC}$ are precomputed, a direct implementation of the estimator \eqref{eq:estimator} has a prohibitive computational scaling ${\cal O}(\ell_{\rm max}^5)$. It is well-known that this issue can be solved when the theoretical bispectrum template can be written in factorized form. While for some bispectrum templates, e.g. local, are explicitly factorizable (see e.g.~ref.~\cite{Komatsu:2003iq}), this is not the case in general. The tensor case considered here, in particular, displays a very complex, nonseparable $\ell$ dependence in $b_{\ell_1 \ell_2 \ell_3}$ (see eq.~\eqref{eq:CMB_bis_general} and refs.~\cite{Shiraishi:2011st,Shiraishi:2012sn,Shiraishi:2012rm,Shiraishi:2013kxa}), so there is no simple solution.

To address this issue, in this paper we rely on the so-called separable modal methodology \cite{Fergusson:2009nv,Fergusson:2010dm}. In this approach, one
considers decompositions of $b_{\ell_1 \ell_2 \ell_3}$ into products of separable modal eigenfunctions in the tetrahedral $\ell$ space. Suitable bases can be found to achieve high correlation between the decomposed and the original shape, using finite, small sets of eigenmodes. This allows the desired construction of approximate separable templates (to arbitrarily high accuracy, at the price of adding more modes). This methodology enables the measurements of many complex shapes, including nonstandard scalar bispectra (e.g., oscillatory bispectra) \cite{Fergusson:2009nv,Fergusson:2010dm,Ade:2013ydc,Fergusson:2014hya,Fergusson:2014tza,Ade:2015ava} and tensor-mode shapes, both in even and odd $\ell_1 + \ell_2 + \ell_3$ domains \cite{Shiraishi:2014roa,Shiraishi:2014ila,Ade:2015ava}.

Following this approach, we decompose our starting template \eqref{eq:CMB_bis_general} as 
\begin{eqnarray}
\frac{v_{\ell_1} v_{\ell_2} v_{\ell_3}}{\sqrt{C_{\ell_1} C_{\ell_2} C_{\ell_3}} } b_{\ell_1 \ell_2 \ell_3}
= \sum_n \alpha_n Q_n(\ell_1, \ell_2, \ell_3)~, \label{eq:Qdec} 
\end{eqnarray}
where the (real) modal basis is given by
\begin{eqnarray}
  Q_{ijk}(\ell_1, \ell_2, \ell_3) = \frac{1}{6} q_{i}(\ell_1) q_{j}(\ell_2) q_{k}(\ell_3) + 5 \, {\rm perms}  
 \equiv
  q_{\{i}(\ell_1) q_{j}(\ell_2) q_{k\}}(\ell_3)~,
\end{eqnarray}
and $v_\ell$ is an arbitrary function introduced to improve convergence by adjusting the overall $\ell$ scaling. For notational convenience, here, the triples $ijk$ in the modal basis are represented with a single index $n$. For the modal elements $q_i(\ell)$, we adopt polynomial eigenfunctions and some specific functions peaking in the squeezed limit, to accelerate convergence in the region where the signal is largest. In the following analysis, we perform the decomposition with $600$ modes, as in ref.~\cite{Ade:2015ava}, achieving more than $90\%$ correlation between the original and the expanded shape.

Factorizing eq.~\eqref{eq:estimator} via eq.~\eqref{eq:Qdec} and the identity
\begin{eqnarray}
\left(
\begin{array}{ccc}
\ell_1 & \ell_2 & \ell_3 \\
m_1 & m_2 & m_3
\end{array}
\right)
h_{\ell_1 \ell_2 \ell_3} = \int d^2 \hat{n}  Y_{\ell_1 m_1}(\hat{n})  Y_{\ell_2 m_2}(\hat{n}) Y_{\ell_3 m_3}(\hat{n})
\end{eqnarray}
yields 
\begin{eqnarray}
\hat{g}_{tss} = \frac{ 1}{N} \sum_n \alpha_{n} \beta_{n} ~, \label{eq:estimator_Q}
\end{eqnarray}
where $\beta_n$ are computed from the products of the maps filtered by $q_i(\ell)$ in $\ell$ space 
\begin{eqnarray}
M_i(\hat{n}) \equiv \sum_{\ell m} q_i(\ell) \frac{a_{\ell m} }{v_{\ell} \sqrt{C_\ell}}  Y_{\ell m}(\hat{n}) ~,
\end{eqnarray}
according to 
\begin{eqnarray}
  \beta_{n \leftrightarrow ijk} \equiv
  \int d^2 \hat{n} 
\left[ M_{\{i}(\hat{n}) M_{j}(\hat{n}) M_{k\}}(\hat{n})
  - 3 \Braket{M_{\{i}(\hat{n}) M_{j}(\hat{n})}_{\rm MC} M_{k\}}(\hat{n}) \right] ~. \label{eq:beta}
\end{eqnarray}
Given that $M_i(\hat{n})$ is precomputed, this step requires ${\cal O}(\ell_{\rm max}^2)$ operations. In the identical manner, the normalization factor also reduces to $N = \sum_{np} \alpha_{n} \gamma_{np} \alpha_{p}$, where
$\gamma_{np} \equiv \Braket{Q_n, Q_p}$ denotes the inner product of the $Q$ basis, computed according to  
\begin{equation}
\Braket{f, g}
\equiv \sum_{\ell_1 \ell_2 \ell_3} 
\left( \frac{h_{\ell_1 \ell_2 \ell_3}}{v_{\ell_1} v_{\ell_2} v_{\ell_3}} \right)^2
f(\ell_1, \ell_2, \ell_3) g(\ell_1, \ell_2, \ell_3)~.
\end{equation}
To obtain $\alpha_n$ from $b_{\ell_1 \ell_2 \ell_3}$, we need to compute another inner product:
\begin{eqnarray}
\alpha_n = \sum_p \gamma_{np}^{-1} 
\Braket{ \frac{v_{\ell_1} v_{\ell_2} v_{\ell_3} b_{\ell_1 \ell_2 \ell_3}}{\sqrt{C_{\ell_1} C_{\ell_2} C_{\ell_3}}}, Q_p(\ell_1,\ell_2,\ell_3) }~. \label{eq:alphaQ}
\end{eqnarray}
The inner product in $\gamma_{np}$ can be always written in separable form, reducing its computational cost to ${\cal O}(\ell_{\rm max})$. There is however no way to factorize the inner product in eq.~\eqref{eq:alphaQ}, since that is where the starting nonseparable shape appears. Computing this product therefore requires ${\cal O}(\ell_{\rm max}^3)$ operations. This is the most time-consuming process in the modal methodology. Note however that this is just a preliminary computation, that needs to be performed only once. Afterwards, the shape is decomposed and only separable quantities enter the actual estimation process. This  makes our bispectrum estimation feasible and fast.

\subsection{Validation tests with non-Gaussian simulations}

Before moving to actual data analysis, we need to perform some validation checks of our estimator, eq.~\eqref{eq:estimator_Q}, for the shape under study. To this end, we estimate $g_{tss}$ from simulated CMB maps including nonzero $g_{tss}$ and check the consistency with input $g_{tss}$.

In the presence of nonzero (but small) $g_{tss}$, the temperature fluctuations can be approximately divided into Gaussian and NG contributions as $a_{\ell m} = a_{\ell m}^{\rm G} + g_{tss} a_{\ell m}^{\rm NG}$. An explicit form of $a_{\ell m}^{\rm NG}$ should be determined as the bispectrum $\Braket{a_{\ell_1 m_1} a_{\ell_2 m_2} a_{\ell_3 m_3}}$ recovers the input theoretical template $g_{tss} B_{\ell_1 \ell_2 \ell_3} \left(
  \begin{array}{ccc}
  \ell_1 & \ell_2 & \ell_3 \\
  m_1 & m_2 & m_3 
  \end{array}
  \right)$, provided that the ${\cal O}(g_{tss}^2)$ contributions and higher-order ones are negligible. A commonly used expression reads
  \begin{equation}
a_{\ell_1 m_1}^{\rm NG} =
\frac{1}{6}
\left[ \prod_{n=2}^3 \sum_{\ell_n m_n} \frac{a_{\ell_n m_n}^{{\rm G}*}}{C_{\ell_n}} \right]
\left(
  \begin{array}{ccc}
  \ell_1 & \ell_2 & \ell_3 \\
  m_1 & m_2 & m_3
  \end{array}
 \right)B_{\ell_1 \ell_2 \ell_3}~. \label{eq:almNG_common}
\end{equation}
  However, ref.~\cite{Hanson:2009kg} has shown that, for the squeezed-type bispectrum, this algorithm can induce spuriously divergent signals on small $\ell_1$ and hence does not work in practice. It was fortunately also shown that this can be cured by using  a slightly different expansion kernel, which in our case reads:
  \begin{equation}
 a_{\ell_1 m_1}^{\rm NG} =
\frac{1}{2}
\left[ \prod_{n=2}^3 \sum_{\ell_n m_n} 
\frac{a_{\ell_n m_n}^{{\rm G}*}}{C_{\ell_n}} \right] 
\left(
  \begin{array}{ccc}
  \ell_1 & \ell_2 & \ell_3 \\
  m_1 & m_2 & m_3
  \end{array}
 \right)
B_{\ell_1 \ell_2 \ell_3}^{(sst)} ~. \label{eq:almNG_sq}
  \end{equation}
  Note that, in the formula above, we have replaced $B_{\ell_1 \ell_2 \ell_3}$ with $B^{(sst)}_{\ell_1 \ell_2 \ell_3}$, i.e. we removed cyclic terms from the original formula, and replaced accordingly the prefactor ${1/6}$ with ${1/2}$. This stabilizes the algorithm against the small-$\ell_1$ divergence, because of the suppression of $B_{\ell_1 \ell_2 \ell_3}^{(sst)}$ outside of the squeezed triangles ($\ell_1 \sim \ell_2 \gg \ell_3$).

 Also for this algorithm, the straightforward computation of eq.~\eqref{eq:almNG_sq} requires ${\cal O}(\ell_{\rm max}^5)$ operations, which are however in principle reduced to ${\cal O}(\ell_{\rm max}^3)$, once separability via modal decomposition is obtained \cite{Fergusson:2009nv,Fergusson:2010dm}. Unfortunately,  expanding $b_{\ell_1 \ell_2 \ell_3}^{(sst)}$ (no cyclic terms) turned out to be very difficult. The asymmetry of this kernel led in fact to very slow convergence. Therefore, we decided in the end to rely on the straightforward, brute-force computation of the nonseparable form \eqref{eq:almNG_sq}. This turned out to be slow but numerically feasible for the angular resolution of this analysis ($\ell < 500$).

\begin{figure}[t]
\begin{center}
    \includegraphics[width=1.\textwidth]{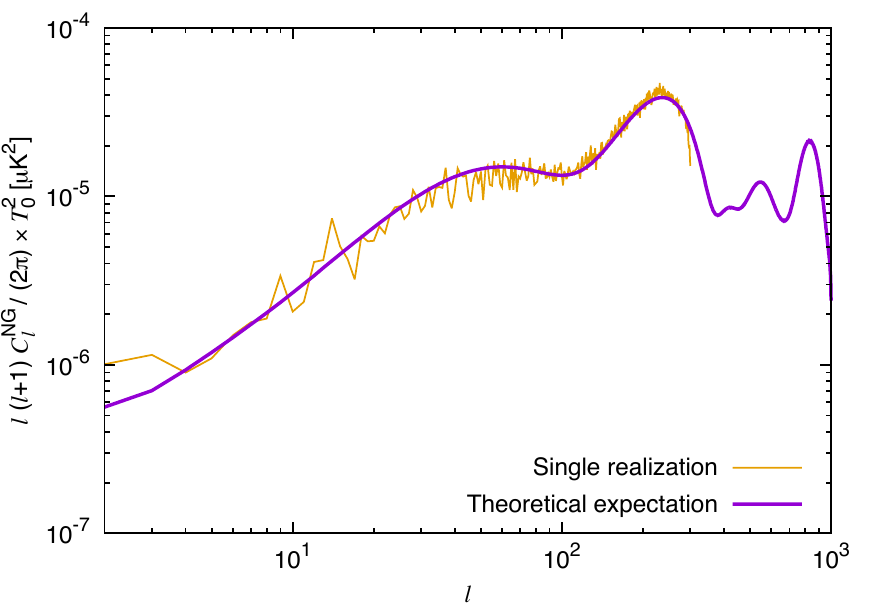}
 \end{center}
    \caption{Angular power spectrum of the NG part of a single realization ($a_{\ell m}^{\rm NG}$), generated from formula~\eqref{eq:almNG_sq}, up to $\ell = 300$ (yellow line), and its theoretical prediction (purple line).} \label{fig:ClNG}
\end{figure}

  To validate our NG estimator, we generate $50$ $a_{\ell m}^{\rm NG}$'s up to $\ell =  300$. The yellow line in figure~\ref{fig:ClNG} corresponds to the angular power spectrum of the NG part of one of the maps. As expected above, there is no pathological enhancement at small $\ell$'s, and the recovered $C_{\ell}$ are fully consistent with their theoretical expectation (purple line in the plot), which reads:
  \begin{eqnarray}
    C_{\ell_1, \rm th}^{\rm NG} =
\frac{1}{4}\sum_{\ell_2 \ell_3}
\frac{
|B_{\ell_1 \ell_2 \ell_3}^{(sst)}|^2 + B_{\ell_1 \ell_2 \ell_3}^{(sst)} B_{\ell_1 \ell_3 \ell_2}^{(sst)} }{(2\ell_1 + 1)C_{\ell_2} C_{\ell_3}}~.
  \end{eqnarray}

\begin{figure}[t]
\begin{center}
    \includegraphics[width=1.\textwidth]{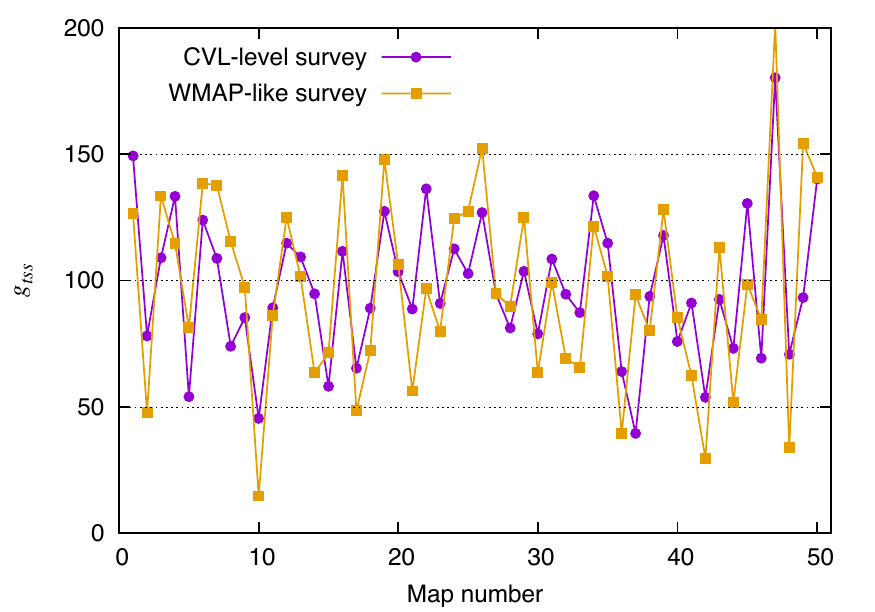}
 \end{center}
    \caption{Map-by-map comparison of estimated $g_{tss}$ assuming a CVL-level full-sky survey (purple line) and the WMAP one (yellow one). The results are obtained from $50$ simulated maps including $g_{tss} = 100$ with $\ell_{\rm max} = 300$.} \label{fig:gtss_sim}
\end{figure}

Having checked the reliability of the NG part of our 50 simulated maps, we now generate 50 NG realizations with $g_{tss} = 100$, i.e. we set $a_{\ell m} = a_{\ell m}^{\rm G} + 100 a_{\ell m}^{\rm NG}$. We then use our estimator~\eqref{eq:estimator_Q} to measure $g_{tss}$, map-by-map. We consider both a CVL-level scenario, with $f_{\rm sky} = 1$, and a realistic one, with the same noise properties and sky coverage as in WMAP data. All CMB maps used for the latter case were processed in accordance with our WMAP data analysis pipeline described at the beginning of this section. Figure~\ref{fig:gtss_sim} shows the map-by-map comparison of estimated $g_{tss}$ for these two cases. It is visually apparent there that the CVL-level case and the WMAP-like one fluctuate around $g_{tss} = 100$ in a very similar way, with the latter case being more scattered than the former, due to the inclusion of mask and noise. The average $g_{tss}$ from the $50$ maps becomes turned out to be $97$ in both cases. The error bar on $g_{tss}$ was derived from 500 Gaussian realizations, obtaining $\Delta g_{tss} = 36$ ($68\%$CL), in excellent agreement with the Fisher matrix forecast displayed in figure~\ref{fig:error_WMAP}. The above results validate our pipeline, showing that the estimator is unbiased and optimal.

\subsection{WMAP limits}

\begin{figure}[t]
\begin{center}
    \includegraphics[width=1.\textwidth]{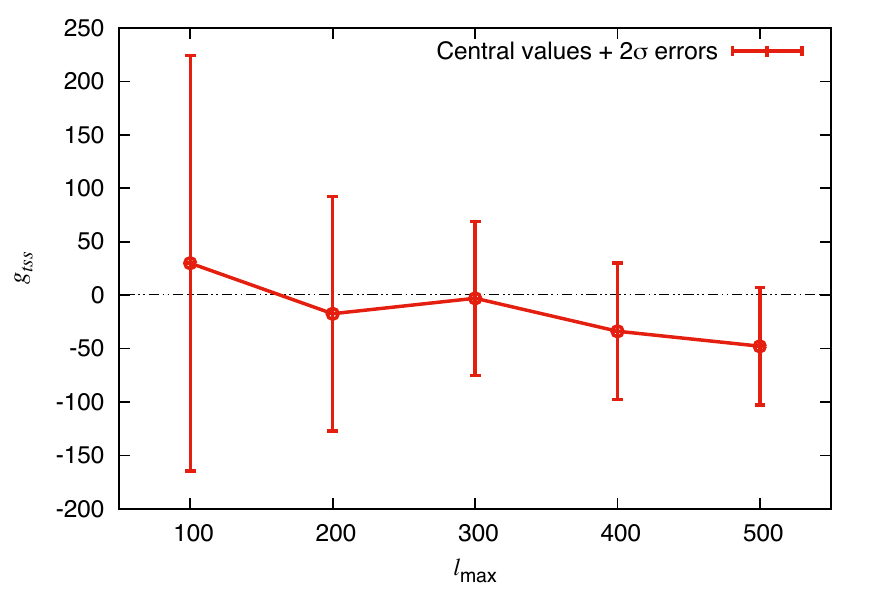}
 \end{center}
    \caption{Central values and $2\sigma$ errors on $g_{tss}$ obtained from the WMAP data as a function of $\ell_{\rm max}$.} \label{fig:gtss_lmax}
\end{figure}

Having tested the estimator on mock realizations, we moved to actual WMAP data and repeated the estimation procedure described above, to find our $g_{tss}$ bounds. We checked stability of the results in the $\ell$-domain by repeating the analysis for varying $\ell_{\rm max}$, from 100 to 500. As a further validation test, we also used the same pipeline to measure the standard $f_{\rm NL}^{\rm local}$ parameter, obtaining fully consistent results with those shown in the literature \cite{Komatsu:2008hk,Smith:2009jr,Komatsu:2010fb,Bennett:2012zja}. Figure~\ref{fig:gtss_lmax} shows the limits on $g_{tss}$ as a function of $\ell_{\rm max}$, indicating no evidence of $g_{tss}$ for $\ell_{\rm max} \leq 500$, at any scale, at $95\%$CL. We take the result at $\ell_{\rm max} = 500$ as our final bound: $g_{tss} = -48 \pm 28$~($68\%$CL).

One may worry here about the contamination due to secondary sources of temperature NG, which we have not considered so far. In particular, it is widely known that the lensed bispectrum can become an important source of bias in the squeezed limit. However, such lensed signal becomes large when higher multipoles are considered, and almost uncorrelated to the primordial one for $\ell_{\rm max} \leq 500$ \cite{Domenech:2017kno}. Therefore, ISW-lensing debiasing is not necessary in our analysis.

\section{Conclusions}\label{sec:conclusions}

In this paper, we have studied the inflationary tensor-scalar-scalar three point function, for models characterized by nonzero graviton mass. A nonvanishing CMB temperature bispectrum is one of the predictions of such models, so we have actually tested it, using WMAP 9-year temperature data. The primordial and the induced CMB bispectrum, $b_{\ell_1 \ell_2 \ell_3}^{(tss)}$, peak in the squeezed limit, in this scenario, and specific Fisher matrix forecasts show that interesting bounds can already be obtained at WMAP angular resolution, which motivated our analysis.

To circumvent nonseparability issues of the primordial shape under study, we have relied on a modal estimation pipeline, which was thoroughly validated on simulated NG maps, generated as a part of this work. After the preliminary validation stage we measured $g_{tss}$ for several $\ell_{\rm max}$'s, finding no evidence for a tensor-scalar-scalar signal at all scales. The most stringent bound is obtained with $\ell_{\rm max} = 500$ and reads $g_{tss} = -48 \pm 28$ ($68\%$CL).

To the best of our knowledge, this is the first paper reporting observational constraints on the primordial tensor-scalar-scalar bispectrum. While we report no evidence of such a signal at the end of our WMAP analysis, it is worth to point out that more sensitive, higher angular resolution surveys, including polarization information, can lead to a further order of magnitude improvement over the current bound \cite{Meerburg:2016ecv,Domenech:2017kno}, thus extending our detectability window. This encourages follow-up investigations with temperature and polarized data measured in {\it Planck} \cite{Ade:2015ava} and forthcoming CMB experiments \cite{Hazumi:2012aa,Abazajian:2016yjj,Finelli:2016cyd}, as part of our future work. An analysis of the {\it Planck} data using the present pipeline, is actually already ongoing within the {\it Planck} collaboration.


\acknowledgments

We thank Paul Shellard for useful discussions. M.\,S. was supported by JSPS Grant-in-Aid for Research Activity Start-up Grant Number 17H07319. M.\,L. acknowledges financial support by ASI Grant 2016-24-H.0. M.\,L acknowledges partial financial support by the ASI/INAF Agreement I/072/09/0 for the Planck LFI Activity of Phase E2. Numerical computations were in part carried out on Cray XC30 at Center for Computational Astrophysics, National Astronomical Observatory of Japan.





\bibliography{paper}

\providecommand{\href}[2]{#2}\begingroup\raggedright\begin{thebibliography}{10}

\bibitem{Acquaviva:2002ud}
V.~Acquaviva, N.~Bartolo, S.~Matarrese and A.~Riotto, \emph{{Second order
  cosmological perturbations from inflation}},
  \href{http://dx.doi.org/10.1016/S0550-3213(03)00550-9}{\emph{Nucl.Phys.} {\bf
  B667} (2003) 119--148}, [\href{http://arxiv.org/abs/astro-ph/0209156}{{\tt
  astro-ph/0209156}}].

\bibitem{Maldacena:2002vr}
J.~M. Maldacena, \emph{{Non-Gaussian features of primordial fluctuations in
  single field inflationary models}},
  \href{http://dx.doi.org/10.1088/1126-6708/2003/05/013}{\emph{JHEP} {\bf 0305}
  (2003) 013}, [\href{http://arxiv.org/abs/astro-ph/0210603}{{\tt
  astro-ph/0210603}}].

\bibitem{Bartolo:2004if}
N.~Bartolo, E.~Komatsu, S.~Matarrese and A.~Riotto, \emph{{Non-Gaussianity from
  inflation: Theory and observations}},
  \href{http://dx.doi.org/10.1016/j.physrep.2004.08.022}{\emph{Phys.Rept.} {\bf
  402} (2004) 103--266}, [\href{http://arxiv.org/abs/astro-ph/0406398}{{\tt
  astro-ph/0406398}}].

\bibitem{Liguori:2010hx}
M.~Liguori, E.~Sefusatti, J.~R. Fergusson and E.~Shellard, \emph{{Primordial
  non-Gaussianity and Bispectrum Measurements in the Cosmic Microwave
  Background and Large-Scale Structure}},
  \href{http://dx.doi.org/10.1155/2010/980523}{\emph{Adv.Astron.} {\bf 2010}
  (2010) 980523}, [\href{http://arxiv.org/abs/1001.4707}{{\tt 1001.4707}}].

\bibitem{Chen:2010xka}
X.~Chen, \emph{{Primordial Non-Gaussianities from Inflation Models}},
  \href{http://dx.doi.org/10.1155/2010/638979}{\emph{Adv. Astron.} {\bf 2010}
  (2010) 638979}, [\href{http://arxiv.org/abs/1002.1416}{{\tt 1002.1416}}].

\bibitem{Komatsu:2010hc}
E.~Komatsu, \emph{{Hunting for Primordial Non-Gaussianity in the Cosmic
  Microwave Background}},
  \href{http://dx.doi.org/10.1088/0264-9381/27/12/124010}{\emph{Class. Quant.
  Grav.} {\bf 27} (2010) 124010}, [\href{http://arxiv.org/abs/1003.6097}{{\tt
  1003.6097}}].

\bibitem{Yadav:2010fz}
A.~P.~S. Yadav and B.~D. Wandelt, \emph{{Primordial Non-Gaussianity in the
  Cosmic Microwave Background}},
  \href{http://dx.doi.org/10.1155/2010/565248}{\emph{Adv. Astron.} {\bf 2010}
  (2010) 565248}, [\href{http://arxiv.org/abs/1006.0275}{{\tt 1006.0275}}].

\bibitem{Bennett:2012zja}
{\scshape WMAP} collaboration, C.~L. Bennett et~al., \emph{{Nine-Year Wilkinson
  Microwave Anisotropy Probe (WMAP) Observations: Final Maps and Results}},
  \href{http://dx.doi.org/10.1088/0067-0049/208/2/20}{\emph{Astrophys. J.
  Suppl.} {\bf 208} (2013) 20}, [\href{http://arxiv.org/abs/1212.5225}{{\tt
  1212.5225}}].

\bibitem{Ade:2013ydc}
{\scshape Planck Collaboration} collaboration, P.~Ade et~al., \emph{{Planck
  2013 Results. XXIV. Constraints on primordial non-Gaussianity}},
  \href{http://arxiv.org/abs/1303.5084}{{\tt 1303.5084}}.

\bibitem{Ade:2015ava}
{\scshape Planck} collaboration, P.~A.~R. Ade et~al., \emph{{Planck 2015
  results. XVII. Constraints on primordial non-Gaussianity}},
  \href{http://dx.doi.org/10.1051/0004-6361/201525836}{\emph{Astron.
  Astrophys.} {\bf 594} (2016) A17},
  [\href{http://arxiv.org/abs/1502.01592}{{\tt 1502.01592}}].

\bibitem{Fergusson:2014hya}
J.~R. Fergusson, H.~F. Gruetjen, E.~P.~S. Shellard and M.~Liguori,
  \emph{{Combining power spectrum and bispectrum measurements to detect
  oscillatory features}},
  \href{http://dx.doi.org/10.1103/PhysRevD.91.023502}{\emph{Phys. Rev.} {\bf
  D91} (2015) 023502}, [\href{http://arxiv.org/abs/1410.5114}{{\tt
  1410.5114}}].

\bibitem{Fergusson:2014tza}
J.~R. Fergusson, H.~F. Gruetjen, E.~P.~S. Shellard and B.~Wallisch,
  \emph{{Polyspectra searches for sharp oscillatory features in cosmic
  microwave sky data}},
  \href{http://dx.doi.org/10.1103/PhysRevD.91.123506}{\emph{Phys. Rev.} {\bf
  D91} (2015) 123506}, [\href{http://arxiv.org/abs/1412.6152}{{\tt
  1412.6152}}].

\bibitem{Shiraishi:2014ila}
M.~Shiraishi, M.~Liguori and J.~R. Fergusson, \emph{{Observed parity-odd CMB
  temperature bispectrum}},
  \href{http://dx.doi.org/10.1088/1475-7516/2015/01/007}{\emph{JCAP} {\bf 1501}
  (2015) 007}, [\href{http://arxiv.org/abs/1409.0265}{{\tt 1409.0265}}].

\bibitem{Shiraishi:2011st}
M.~Shiraishi, D.~Nitta and S.~Yokoyama, \emph{{Parity Violation of Gravitons in
  the CMB Bispectrum}},
  \href{http://dx.doi.org/10.1143/PTP.126.937}{\emph{Prog.Theor.Phys.} {\bf
  126} (2011) 937--959}, [\href{http://arxiv.org/abs/1108.0175}{{\tt
  1108.0175}}].

\bibitem{Shiraishi:2013vha}
M.~Shiraishi, \emph{{Polarization bispectrum for measuring primordial magnetic
  fields}}, \href{http://dx.doi.org/10.1088/1475-7516/2013/11/006}{\emph{JCAP}
  {\bf 1311} (2013) 006}, [\href{http://arxiv.org/abs/1308.2531}{{\tt
  1308.2531}}].

\bibitem{Shiraishi:2013kxa}
M.~Shiraishi, A.~Ricciardone and S.~Saga, \emph{{Parity violation in the CMB
  bispectrum by a rolling pseudoscalar}},
  \href{http://dx.doi.org/10.1088/1475-7516/2013/11/051}{\emph{JCAP} {\bf 1311}
  (2013) 051}, [\href{http://arxiv.org/abs/1308.6769}{{\tt 1308.6769}}].

\bibitem{Shiraishi:2016yun}
M.~Shiraishi, C.~Hikage, R.~Namba, T.~Namikawa and M.~Hazumi, \emph{{Testing
  statistics of the CMB B-mode polarization toward unambiguously establishing
  quantum fluctuation of the vacuum}},
  \href{http://dx.doi.org/10.1103/PhysRevD.94.043506}{\emph{Phys. Rev.} {\bf
  D94} (2016) 043506}, [\href{http://arxiv.org/abs/1606.06082}{{\tt
  1606.06082}}].

\bibitem{Tahara:2017wud}
H.~W.~H. Tahara and J.~Yokoyama, \emph{{CMB B-mode auto-bispectrum produced by
  primordial gravitational waves}},
  \href{http://arxiv.org/abs/1704.08904}{{\tt 1704.08904}}.

\bibitem{Kamionkowski:2010rb}
M.~Kamionkowski and T.~Souradeep, \emph{{The Odd-Parity CMB Bispectrum}},
  \href{http://dx.doi.org/10.1103/PhysRevD.83.027301}{\emph{Phys.Rev.} {\bf
  D83} (2011) 027301}, [\href{http://arxiv.org/abs/1010.4304}{{\tt
  1010.4304}}].

\bibitem{Shiraishi:2012sn}
M.~Shiraishi, \emph{{Parity violation of primordial magnetic fields in the CMB
  bispectrum}},
  \href{http://dx.doi.org/10.1088/1475-7516/2012/06/015}{\emph{JCAP} {\bf 1206}
  (2012) 015}, [\href{http://arxiv.org/abs/1202.2847}{{\tt 1202.2847}}].

\bibitem{Shiraishi:2014roa}
M.~Shiraishi, M.~Liguori and J.~R. Fergusson, \emph{{General parity-odd CMB
  bispectrum estimation}},
  \href{http://dx.doi.org/10.1088/1475-7516/2014/05/008}{\emph{JCAP} {\bf 1405}
  (2014) 008}, [\href{http://arxiv.org/abs/1403.4222}{{\tt 1403.4222}}].

\bibitem{Domenech:2017kno}
G.~Dom\'enech, T.~Hiramatsu, C.~Lin, M.~Sasaki, M.~Shiraishi and Y.~Wang,
  \emph{{CMB Scale Dependent Non-Gaussianity from Massive Gravity during
  Inflation}},
  \href{http://dx.doi.org/10.1088/1475-7516/2017/05/034}{\emph{JCAP} {\bf 1705}
  (2017) 034}, [\href{http://arxiv.org/abs/1701.05554}{{\tt 1701.05554}}].

\bibitem{Namba:2015gja}
R.~Namba, M.~Peloso, M.~Shiraishi, L.~Sorbo and C.~Unal, \emph{{Scale-dependent
  gravitational waves from a rolling axion}},
  \href{http://dx.doi.org/10.1088/1475-7516/2016/01/041}{\emph{JCAP} {\bf 1601}
  (2016) 041}, [\href{http://arxiv.org/abs/1509.07521}{{\tt 1509.07521}}].

\bibitem{Agrawal:2017awz}
A.~Agrawal, T.~Fujita and E.~Komatsu, \emph{{Large Tensor Non-Gaussianity from
  Axion-Gauge Fields Dynamics}},  \href{http://arxiv.org/abs/1707.03023}{{\tt
  1707.03023}}.

\bibitem{Hinshaw:2012aka}
{\scshape WMAP} collaboration, G.~Hinshaw et~al., \emph{{Nine-Year Wilkinson
  Microwave Anisotropy Probe (WMAP) Observations: Cosmological Parameter
  Results}},
  \href{http://dx.doi.org/10.1088/0067-0049/208/2/19}{\emph{Astrophys. J.
  Suppl.} {\bf 208} (2013) 19}, [\href{http://arxiv.org/abs/1212.5226}{{\tt
  1212.5226}}].

\bibitem{Shiraishi:2010kd}
M.~Shiraishi, D.~Nitta, S.~Yokoyama, K.~Ichiki and K.~Takahashi, \emph{{CMB
  Bispectrum from Primordial Scalar, Vector and Tensor non-Gaussianities}},
  \href{http://dx.doi.org/10.1143/PTP.125.795}{\emph{Prog.Theor.Phys.} {\bf
  125} (2011) 795--813}, [\href{http://arxiv.org/abs/1012.1079}{{\tt
  1012.1079}}].

\bibitem{Fergusson:2009nv}
J.~R. Fergusson, M.~Liguori and E.~P.~S. Shellard, \emph{{General CMB and
  Primordial Bispectrum Estimation I: Mode Expansion, Map-Making and Measures
  of $f_{\rm NL}$}},
  \href{http://dx.doi.org/10.1103/PhysRevD.82.023502}{\emph{Phys. Rev.} {\bf
  D82} (2010) 023502}, [\href{http://arxiv.org/abs/0912.5516}{{\tt
  0912.5516}}].

\bibitem{Fergusson:2010dm}
J.~R. Fergusson, M.~Liguori and E.~P.~S. Shellard, \emph{{The CMB Bispectrum}},
  \href{http://dx.doi.org/10.1088/1475-7516/2012/12/032}{\emph{JCAP} {\bf 1212}
  (2012) 032}, [\href{http://arxiv.org/abs/1006.1642}{{\tt 1006.1642}}].

\bibitem{Meerburg:2016ecv}
P.~D. Meerburg, J.~Meyers, A.~van Engelen and Y.~Ali-Haïmoud, \emph{{CMB
  B-mode non-Gaussianity}},
  \href{http://dx.doi.org/10.1103/PhysRevD.93.123511}{\emph{Phys. Rev.} {\bf
  D93} (2016) 123511}, [\href{http://arxiv.org/abs/1603.02243}{{\tt
  1603.02243}}].

\bibitem{Komatsu:2001rj}
E.~Komatsu and D.~N. Spergel, \emph{{Acoustic signatures in the primary
  microwave background bispectrum}},
  \href{http://dx.doi.org/10.1103/PhysRevD.63.063002}{\emph{Phys. Rev.} {\bf
  D63} (2001) 063002}, [\href{http://arxiv.org/abs/astro-ph/0005036}{{\tt
  astro-ph/0005036}}].

\bibitem{Smith:2007rg}
K.~M. Smith, O.~Zahn and O.~Dore, \emph{{Detection of Gravitational Lensing in
  the Cosmic Microwave Background}},
  \href{http://dx.doi.org/10.1103/PhysRevD.76.043510}{\emph{Phys. Rev.} {\bf
  D76} (2007) 043510}, [\href{http://arxiv.org/abs/0705.3980}{{\tt
  0705.3980}}].

\bibitem{Senatore:2009gt}
L.~Senatore, K.~M. Smith and M.~Zaldarriaga, \emph{{Non-Gaussianities in Single
  Field Inflation and their Optimal Limits from the WMAP 5-year Data}},
  \href{http://dx.doi.org/10.1088/1475-7516/2010/01/028}{\emph{JCAP} {\bf 1001}
  (2010) 028}, [\href{http://arxiv.org/abs/0905.3746}{{\tt 0905.3746}}].

\bibitem{Elsner:2012fe}
F.~Elsner and B.~D. Wandelt, \emph{{Efficient Wiener filtering without
  preconditioning}},
  \href{http://dx.doi.org/10.1051/0004-6361/201220586}{\emph{Astron.
  Astrophys.} {\bf 549} (2013) A111},
  [\href{http://arxiv.org/abs/1210.4931}{{\tt 1210.4931}}].

\bibitem{Gruetjen:2015sta}
H.~F. Gruetjen, J.~R. Fergusson, M.~Liguori and E.~P.~S. Shellard, \emph{{Using
  inpainting to construct accurate cut-sky CMB estimators}},
  \href{http://dx.doi.org/10.1103/PhysRevD.95.043532}{\emph{Phys. Rev.} {\bf
  D95} (2017) 043532}, [\href{http://arxiv.org/abs/1510.03103}{{\tt
  1510.03103}}].

\bibitem{Bucher:2015ura}
M.~Bucher, B.~Racine and B.~van Tent, \emph{{The binned bispectrum estimator:
  template-based and non-parametric CMB non-Gaussianity searches}},
  \href{http://dx.doi.org/10.1088/1475-7516/2016/05/055}{\emph{JCAP} {\bf 1605}
  (2016) 055}, [\href{http://arxiv.org/abs/1509.08107}{{\tt 1509.08107}}].

\bibitem{Komatsu:2008hk}
{\scshape WMAP Collaboration} collaboration, E.~Komatsu et~al.,
  \emph{{Five-Year Wilkinson Microwave Anisotropy Probe (WMAP) Observations:
  Cosmological Interpretation}},
  \href{http://dx.doi.org/10.1088/0067-0049/180/2/330}{\emph{Astrophys.J.Suppl.}
  {\bf 180} (2009) 330--376}, [\href{http://arxiv.org/abs/0803.0547}{{\tt
  0803.0547}}].

\bibitem{Lambda}
\url{http://lambda.gsfc.nasa.gov}.

\bibitem{Komatsu:2003iq}
E.~Komatsu, D.~N. Spergel and B.~D. Wandelt, \emph{{Measuring primordial
  non-Gaussianity in the cosmic microwave background}},
  \href{http://dx.doi.org/10.1086/491724}{\emph{Astrophys.J.} {\bf 634} (2005)
  14--19}, [\href{http://arxiv.org/abs/astro-ph/0305189}{{\tt
  astro-ph/0305189}}].

\bibitem{Shiraishi:2012rm}
M.~Shiraishi, D.~Nitta, S.~Yokoyama and K.~Ichiki, \emph{{Optimal limits on
  primordial magnetic fields from CMB temperature bispectrum of passive
  modes}}, \href{http://dx.doi.org/10.1088/1475-7516/2012/03/041}{\emph{JCAP}
  {\bf 1203} (2012) 041}, [\href{http://arxiv.org/abs/1201.0376}{{\tt
  1201.0376}}].

\bibitem{Hanson:2009kg}
D.~Hanson, K.~M. Smith, A.~Challinor and M.~Liguori, \emph{{CMB lensing and
  primordial non-Gaussianity}},
  \href{http://dx.doi.org/10.1103/PhysRevD.80.083004}{\emph{Phys. Rev.} {\bf
  D80} (2009) 083004}, [\href{http://arxiv.org/abs/0905.4732}{{\tt
  0905.4732}}].

\bibitem{Smith:2009jr}
K.~M. Smith, L.~Senatore and M.~Zaldarriaga, \emph{{Optimal limits on
  $f_{NL}^{local}$ from WMAP 5-year data}},
  \href{http://dx.doi.org/10.1088/1475-7516/2009/09/006}{\emph{JCAP} {\bf 0909}
  (2009) 006}, [\href{http://arxiv.org/abs/0901.2572}{{\tt 0901.2572}}].

\bibitem{Komatsu:2010fb}
{\scshape WMAP Collaboration} collaboration, E.~Komatsu et~al.,
  \emph{{Seven-Year Wilkinson Microwave Anisotropy Probe (WMAP) Observations:
  Cosmological Interpretation}},
  \href{http://dx.doi.org/10.1088/0067-0049/192/2/18}{\emph{Astrophys.J.Suppl.}
  {\bf 192} (2011) 18}, [\href{http://arxiv.org/abs/1001.4538}{{\tt
  1001.4538}}].

\bibitem{Hazumi:2012aa}
{\scshape LiteBIRD} collaboration, M.~Hazumi et~al., \emph{{LiteBIRD: a small
  satellite for the study of B-mode polarization and inflation from cosmic
  background radiation detection}},
  \href{http://dx.doi.org/10.1117/12.926743}{\emph{Proc. SPIE Int. Soc. Opt.
  Eng.} {\bf 8442} (2012) 844219}.

\bibitem{Abazajian:2016yjj}
{\scshape CMB-S4} collaboration, K.~N. Abazajian et~al., \emph{{CMB-S4 Science
  Book, First Edition}},  \href{http://arxiv.org/abs/1610.02743}{{\tt
  1610.02743}}.

\bibitem{Finelli:2016cyd}
{\scshape CORE} collaboration, F.~Finelli et~al., \emph{{Exploring Cosmic
  Origins with CORE: Inflation}},  \href{http://arxiv.org/abs/1612.08270}{{\tt
  1612.08270}}.

\end{thebibliography}\endgroup
\end{document}